\documentclass[prb,twocolumn,nofootinbib,amsmath,amssymb,longbibliography]{revtex4-1}
\usepackage{graphicx}
\usepackage{float}
\usepackage[usenames,dvipsnames]{xcolor}
\usepackage{pdfpages}

\makeatletter
\AtBeginDocument{\let\LS@rot\@undefined}
\makeatother

\def\supplementfilename{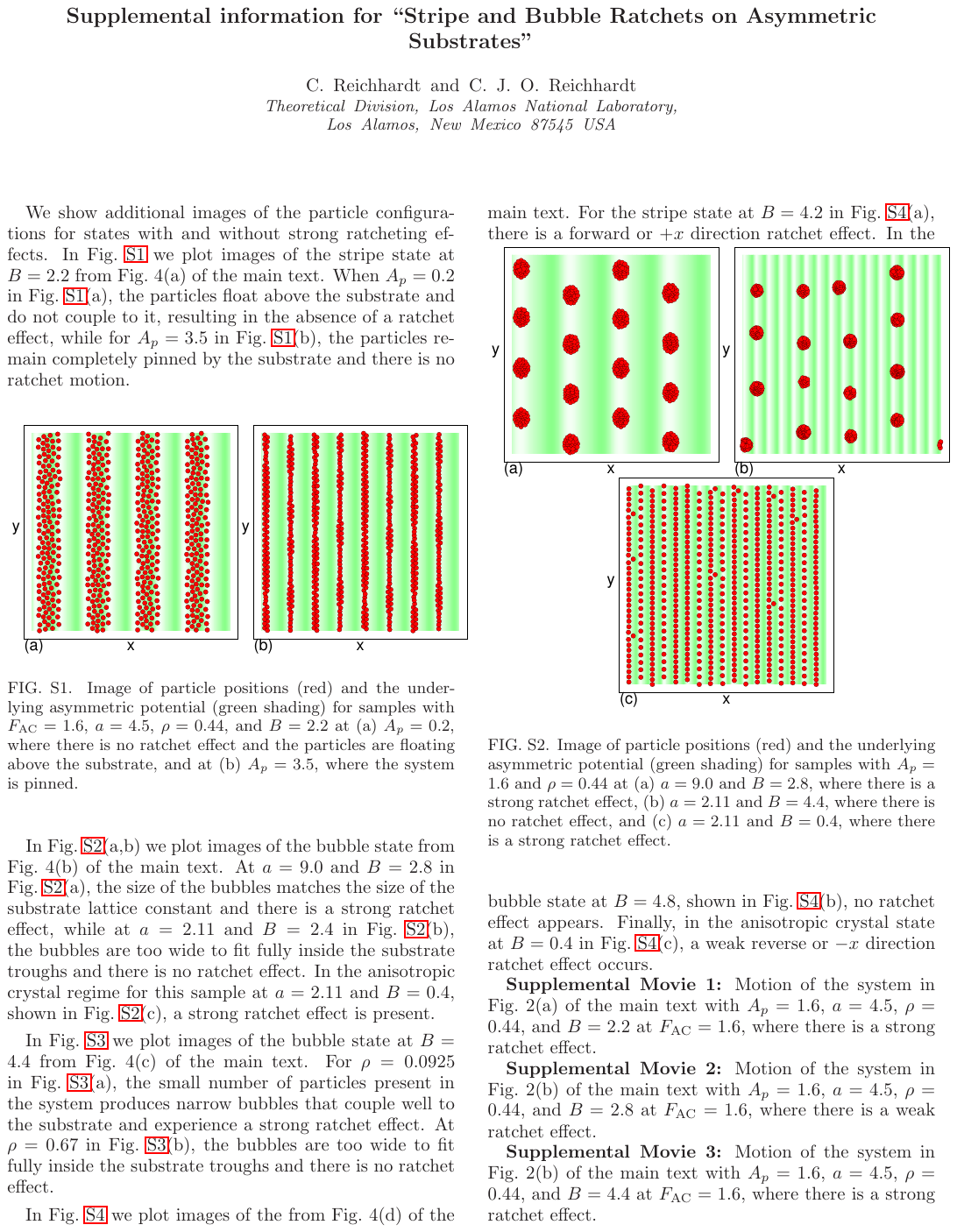}
\pdfximage{\supplementfilename}

\begin{document}

\title{
Stripe and Bubble Ratchets on Asymmetric Substrates 
}
\author{
C. Reichhardt and C. J. O. Reichhardt 
} 
\address{
Theoretical Division and Center for Nonlinear Studies,
Los Alamos National Laboratory, Los Alamos, New Mexico 87545, USA
}

\begin{abstract}
We show that a variety of non-monotonic ratchet effects can arise when mesophase pattern-forming systems, which exhibit anisotropic crystal, stripe, and bubble reigmes, are coupled to one-dimensional asymmetric substrates under ac driving. The ratchet efficiency and direction of motion are determined by how well the mesophase morphology matches the periodicity and shape of the substrate. Stripe states that are aligned with the substrate show the strongest ratchet effect, large bubbles show a weak ratchet effect, and small bubbles show a strong ratchet effect with an efficiency that oscillates as a function of ac drive amplitude. We map out the different rectification phases as a function of the pattern morphology, substrate strength, and ac drive amplitude. The pronounced ratchet effects that we observe in some regimes can be exploited for pattern sorting in hard and soft matter systems.
\end{abstract}

\maketitle

{\it Introduction---}
In a rocking ratchet, ac driving produces
a net dc drift of particles that are
coupled to an underlying asymmetric substrate
\cite{Magnasco93,Faucheux95,Reimann02}.
Ratchet effects can also occur for
Brownian particles moving over asymmetric substrates that flash on and off
cyclically \cite{Astumian94,Rousselet94}.
Such effects have been studied extensively
in a wide range of particle-like systems, including
soft matter \cite{Rousselet94,Marquet02}, biological
systems \cite{Lau17}, and
active matter \cite{Reichhardt17a}.
Ratchet effects also arise in numerous
hard condensed matter systems such as
vortices in type-II superconductors \cite{Lee99,Lu07},
magnetic domain walls \cite{Franken12},
skyrmions \cite{Reichhardt15a},
Wigner crystals \cite{Reichhardt23a},
cold atoms \cite{MenneratRobilliard99,Salger09},
and quantum systems \cite{Linke99,Grossert16}.

Most studies of ratchet effects have focused on point-like particles or
particles that interact via excluded volume;
however, there is a broad class of systems that form
mesophase or patterned states
with higher-order structure
such as stripes, labyrinths, void lattices,
and bubbles \cite{Seul95,Harrison00,Stoycheva00,Nelissen05,Imperio06}.
This type of pattern formation can occur when the particle-particle
interaction potential contains multiple length scales
\cite{Malescio03,Glaser07,Edlund10}
or when there is a competition between
short range attractive and longer range repulsive
interactions
\cite{Stoycheva00,Reichhardt03a,Sciortino04,Nelissen05,Liu08,Reichhardt10,Chen11,McDermott14,Hooshanginejad24}.
Mesophases appear in soft matter systems for certain types
of colloidal particles, emulsions, binary fluids,
and block-copolymer systems,
as well as in hard matter systems
where bubble and stripe phases are observed
for two-dimensional electrons in magnetic fields
\cite{Fogler96,Moessner96,Fradkin99,Lilly99,Gores07},
certain types of
superconducting vortex states
\cite{Babaev05,Xu11,Lin11,Zhao12,Sellin13,Meng17},
charge ordering systems \cite{Emery99,Stojkovic00},
and magnetic textures such as stripes and skyrmion bubbles \cite{Reichhardt24}.

If a mesoscale pattern forming state is coupled to a periodic
asymmetric substrate and subjected to ac driving,
it should be possible to realize a ratchet effect.
It has not previously been explored
whether ratchet effects occur in such systems or,
if so, what type of ratchet effects are present.
New ratchet phenomena could arise when the periodicity or morphology of
the pattern matches the substrate periodicity.
There have been studies of Leidenfrost bubbles coupled to asymmetric substrates
where the bubble widths can span the widths of several substrate minima
\cite{Linke06, Lagubeau11},
but the mechanism for the motion is very different from that
in particle based systems under ac driving.

In this work, we consider an assembly of particles that have competing
long-range repulsion and short-range attraction where we vary
the relative strength
of the short range attraction.
This model has previously been shown
to support a variety of pattern forming states,
including anisotropic crystals, stripes, and bubbles
\cite{Reichhardt03a,Nelissen05,Reichhardt10,Hooshanginejad24}.
When the pattern forming states are coupled to an asymmetric substrate
and subjected to ac driving,
we find a strongly non-monotonic ratchet behavior that depends on the
morphology of the pattern.
The stripe or highly anisotropic crystalline states
can align with the quasi-one-dimensional substrate and
generally exhibit a pronounced ratchet effect.
Just above the stripe-to-bubble transition,
the bubbles are wider than the substrate spacing, can easily
distort, and can break up during an ac driving cycle,
and in this regime,
the ratchet effect is heavily reduced or absent.
Small bubbles undergo
a strong ratchet effect when the bubble diameter
is smaller than the substrate lattice constant and the bubbles are
stiff enough to remain intact across the entire ac driving cycle.
The behavior of very small bubbles is identical to that of
a single particle on an asymmetric substrate,
and shows an
oscillating ratchet efficiency as a function of
ac drive amplitude.
We demonstrate that reversals in the
direction of ratcheting motion can occur as a function of system density,
and show that the ratchet effects are robust
over a range of interactions, ac drive amplitudes, and
particle densities.
The ratchet effect we observe could be used to detect
different patterns or as a method for pattern sorting.

{\it Simulation---}
We consider a two-dimensional system
of size $L \times L$ with $L=36$
that contains $N$ particles
with a pairwise interaction potential that has
a competition between long range repulsion and short-range attraction,
\begin{equation}
V(R_{ij}) = \frac{1}{R_{ij}} - B\exp(-\kappa R_{ij}) \ .
\end{equation}
The distance between particle $i$ and $j$ is
$R_{ij}=|{\bf R}_i-{\bf R}_j|$,
and the first term on the right hand side
is the long-range repulsive Coulomb term, which favors
formation of a uniform triangular lattice.
The second term is a short-range attraction that favors
clump formation.
The strength and range of the attractive term is controlled by
the values of
$B$ and $\kappa$, and here we fix $\kappa=1.0$ and vary $B$.
At small $R_{ij}$, the repulsive term
dominates, so complete collapse of the particles onto a point
does not occur.
The overall potential is repulsive at long length scales, attractive at
intermediate scales, and repulsive again at short length scales.
In the absence of substrate, as $B$ increases, the system passes through
a series of states consisting of
uniform crystal, anisotropic crystal, stripes, large bubbles, and small
bubbles \cite{Reichhardt03a,Nelissen05,Reichhardt10,Xu21}.
The particle density is given by $\rho=N/L^2$.
Recently, this interaction potential was
used to study dc depinning on a symmetric one-dimensional substrate,
where it was shown 
that the stripes are strongly pinned and
the bubbles have a non-monotonic depinning threshold \cite{Reichhardt24}.

The dynamics of particle $i$ in a sample with
periodic boundary conditions in the $x$ and $y$-directions is obtained
using the following
overdamped equation:
\begin{equation}
\eta \frac{d {\bf R}_{i}}{dt} =
-\sum^{N}_{j \neq i} \nabla V(R_{ij}) + {\bf F}^{s}_{i} +
        {\bf F}_{\rm AC} ,
\end{equation}
where the damping term $\eta$ is set to $\eta=1.0$.
The first term on the right
is the particle-particle interaction force,
the second term is the substrate force 
${\bf F}^{s}_{i}=\nabla U(x_i)$, and ${\bf F}_{\rm AC}$ is the ac driving force.
The substrate potential has the form
\begin{equation}
U(x_i) = -U_{p}[\sin(2\pi x_i/a) + 0.25\sin(4\pi x_i/a)] \ ,
\end{equation}
where $x_i$ is the $x$ coordinate of particle $i$
and $a$ is the substrate lattice constant.
This potential has been used previously to study ratchet effects in
soft and hard matter systems \cite{Lee99,Reimann02,Reichhardt15a,Reichhardt23a}.
We characterize the substrate strength by $A_{p} = U_{p}/2\pi$,
so that the maximum pinning force in the hard
or $-x$ direction 
is $F^{\rm hard}_{p} =1.5A_p$ and in the easy or $+x$ direction is
$F^{\rm easy}_{p} = 0.725A_{p}$.
The ac driving force has the form
${\bf F}_{\rm AC}= F_{\rm AC}\sin(\omega t)$, where 
we fix $\omega=0.19$.
For each set of parameters, we wait
$10^5$ simulation time steps after starting the simulation
to avoid any transient effects, and then measure the
particle velocity in the direction of the drive
averaged over 100 ac drive cycles,
$\langle V\rangle = \sum^{N}_i{\bf v}_i\cdot {\hat {\bf x}}$.

\begin{figure}
  \centering
\includegraphics[width=\columnwidth]{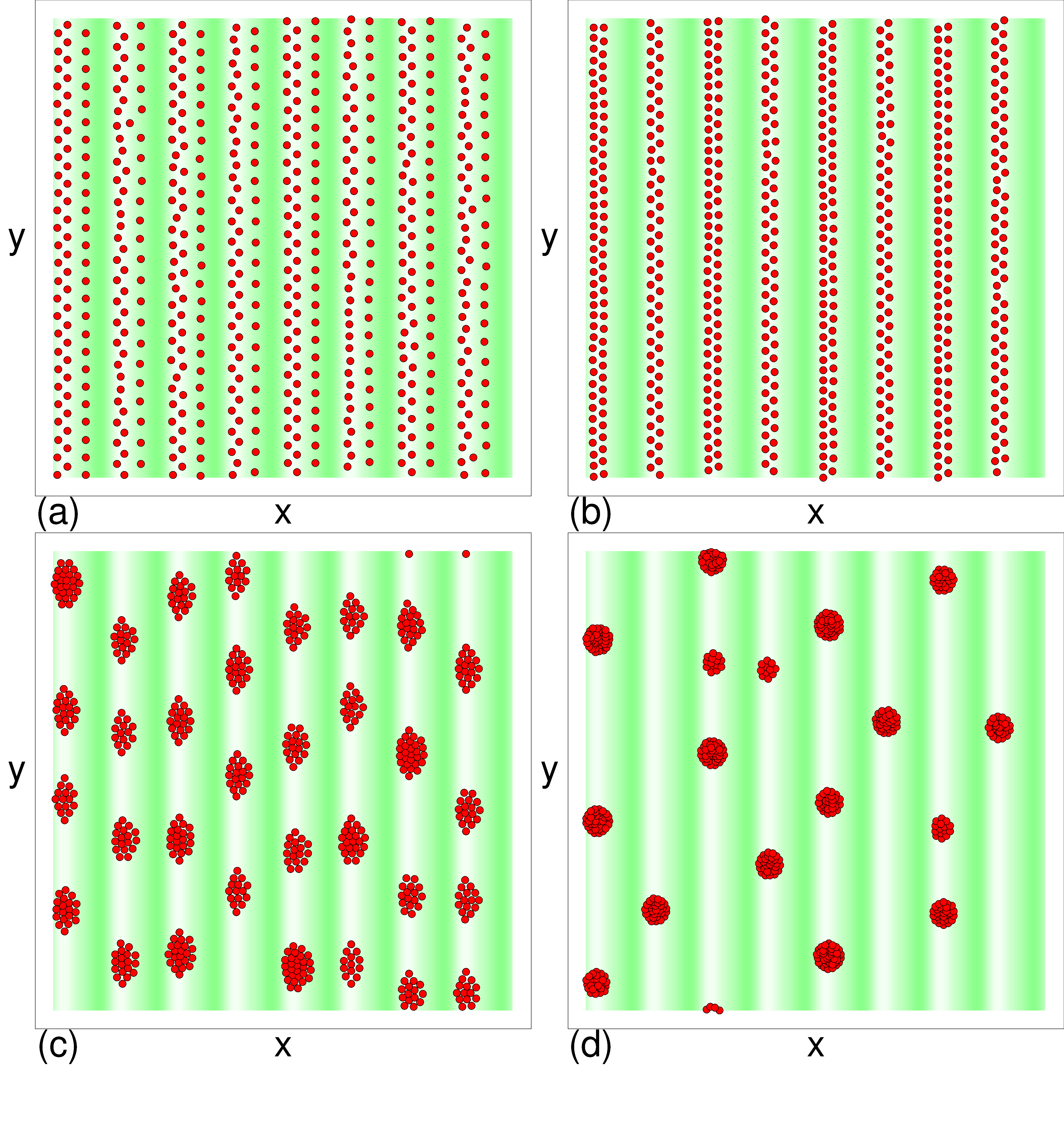}
\caption{
Image of particle positions (red) and the underlying asymmetric potential
(green shading) for samples with
substrate strength $A_{p} = 1.6$,
substrate lattice constant $a=4.5$, and particle density $\rho = 0.44$
under zero driving, $F_{\rm AC}=0$.
(a) $B = 0.0$, where the interaction potential is purely repulsive
and the system forms an anisotropic crystal.
(b) An aligned stripe state at $B = 2.2$.
(c) Disordered large bubbles at $B  =  2.6$.
(d) Small circular bubbles at $ B = 4.4$.
} 
\label{fig:1} 
\end{figure}

{\it Results---}
In Fig.~\ref{fig:1} we show images of the particle configurations
under zero drive, $F_{\rm AC}=0$, in a system with
a substrate strength of $A_{p} = 1.6$, a particle density of $\rho = 0.44$,
and a substrate lattice spacing of $a = 4.5$.
Figure~\ref{fig:1}(a) shows the configuration
at $B = 0.0$ where the
particle-particle interactions are purely repulsive
and an anisotropic crystal appears due to the influence of
the substrate.
At $B=2.2$ in Fig.~\ref{fig:1}(b),
we find an aligned stripe state
with two rows of particles per substrate minimum.
In the bubble state at $B=2.6$ shown in
Fig.~\ref{fig:1}(c),
the bubbles exhibit an asymmetric distortion
due to the substrate.
For $B=4.4$ in Fig.~\ref{fig:1}(d),
the bubbles are much smaller
and have a circular shape.
For this choice of $A_p$ and $\rho$,
the system forms an anisotropic crystal for
$B < 1.8$, a stripe state for $1.8 \leq B \leq 2.4$, and bubbles for $B > 2.4$.

\begin{figure}
  \centering
\includegraphics[width=\columnwidth]{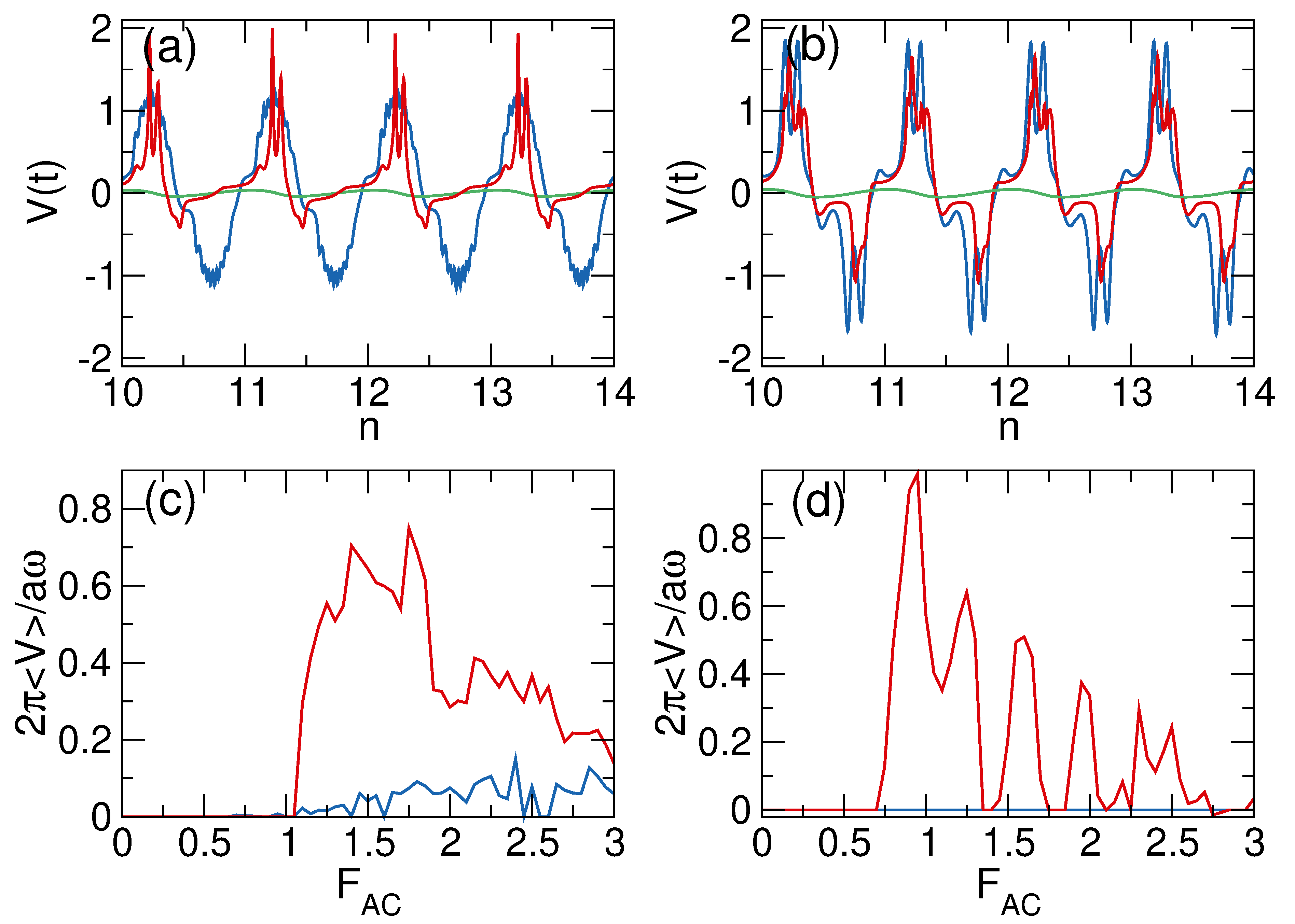}
\caption{
(a) The instantaneous velocity $V(t)$ vs time in number of ac drive cycles $n$
for the system in Fig.~\ref{fig:1} with $A_p=1.6$, $a=4.5$, and
$\rho=0.44$ for different combinations of $B$ and $F_{\rm AC}$.
Blue: $B = 0.0$ and $F_{\rm AC} = 1.6$.
Red: $B = 2.2$ and $F_{\rm AC} = 1.6$.
Green: $B = 2.2$ and $F_{\rm AC} = 0.6$.
(b) The same as in (a).
Blue: $B = 2.8$ and $F_{\rm AC} = 1.6$.
Red: $B = 4.4$ and $F_{\rm AC} = 1.6$.
Green: $B = 4.4$ and $F_{\rm AC} = 0.6$.
Animations of some of these
states appear in the Supplemental Material
\cite{Suppl}.
(c) The normalized velocity
$2\pi\langle V\rangle/\omega a$ vs $F_{\rm AC}$
for a system with $A_p=1.0$, $a=4.5$, and $\rho=0.44$
at
$B = 0.0$ (blue) and $B = 2.2$ (red).
(d) The same for $B = 2.8$ (blue) and $B = 4.4$ (red).
}
\label{fig:2}
\end{figure} 

We next consider the effect of applying an ac drive.
In Fig.~\ref{fig:2}(a)
we plot a time series of the
instantaneous particle velocity $V(t)$  versus ac cycle number $n$ for
the system from Fig.~\ref{fig:1} under different combinations
of $B$ and $F_{\rm AC}$.
When $F_{\rm AC}=1.6$, 
at $B=0.0$ the response in the $+x$ and $-x$ directions
is mostly symmetric, while
at $B = 2.2$ in the stripe state,
the velocity is strongly asymmetric, indicating
the presence of a ratchet effect.
When $F_{\rm AC}$ is reduced to $F_{\rm AC}=0.5$ for $B=2.2$,
the response is symmetric again
and the particles simply oscillate within the substrate minima.
Figure~\ref{fig:2}(b) shows $V(t)$ curves obtained in the
bubble state.
When $F_{\rm AC}=1.6$, at $B=2.8$ the response is largely symmetric so
the ratchet effect is minimized,
while at $B=4.4$,
the response is more asymmetric since there are two velocity peaks in the
$+x$ direction for every individual velocity peak in the $-x$
direction, resulting in an enhanced ratchet effect.
For $F_{\rm AC}=0.5$ and $B=4.4$,
the system is pinned.
These results indicate that the ratchet response in the
bubble state is affected by the size of the bubbles.

To further illustrate the differences in the ratchet behaviors,
in Fig.~\ref{fig:2}(c) we plot the scaled average velocity
$2\pi\langle V\rangle/a\omega$ versus $F_{\rm AC}$ in a system
with $A_p=1.0$.
Here, a value of
$2\pi\langle V\rangle/a\omega=1.0$ indicates that the particles
translate an average of one substrate lattice constant per ac
drive cycle.
When $B = 0.0$,
there is a weak ratchet effect at larger values of $F_{\rm AC}$,
while the stripe state at
$B  = 2.2$ 
shows a strong ratchet effect for
$F_{\rm AC} > 1.0$.
The plots of $2\pi\langle V\rangle/a\omega$ 
versus $F_{\rm AC}$
in Fig.~\ref{fig:2}(d) indicate that for $B = 2.8$,
there is no ratchet effect at any ac drive amplitude,
while at $B = 4.4$, there is
a strong ratchet effect as well as an oscillation of the ratchet
efficiency as function of $F_{\rm AC}$. 
The oscillatory ratchet behavior arises from
a resonance effect with the substrate period.
At the first peak in $2\pi\langle V\rangle/a\omega$, which
falls at $F_{\rm AC} = 0.95$,
the bubbles translate by one substrate
lattice constant during each ac drive cycle,
while the peaks at higher values of $F_{\rm AC}$ correspond to
ratchet states in which
the bubbles traverse multiple substrate minima during each ac drive
cycle but have a net translation of less than one substrate lattice
constant.
Similar oscillatory ratchet behavior
has been observed previously for individual particles
ratcheting over
asymmetric substrates \cite{Lee99,Reichhardt15a}.
Increasing the strength of the collective interactions between
the particles washes out the resonances with the substrate.

The non-monotonic efficiency of the ratchet effect as a function of $B$
results from the differing ability of the distinct pattern morphologies
to couple to the substrate.
For the anisotropic crystals at low $B$,
the repulsive interaction term dominates and the
particles try to spread apart as shown 
in Fig.~\ref{fig:1}(a), so some of the particles are unable to sit in
the substrate minima and the ratchet effect is reduced.
The stripe state illustrated in Fig.~\ref{fig:1}(b) easily aligns with
the substrate,
so a majority of the particles can reach the substrate minima and
a strong ratchet effect appears.
When large bubbles are present, as in Fig.~\ref{fig:1}(c),
a portion of the particles cannot fit inside the substrate minima
since the bubble diameter is too wide,
which reduces the coupling to the substrate and thereby reduces the
magnitude of the ratchet effect; however,
the small bubbles that form
at higher $B$ fit more easily into the poetical minima,
as shown in Fig.~\ref{fig:1}(d), and the ratchet effect becomes large.
For very large $B$, not only are the bubbles smaller
but also there are fewer bubbles,
so the system approaches the single particle ratchet response limit.

\begin{figure}
  \centering
\includegraphics[width=\columnwidth]{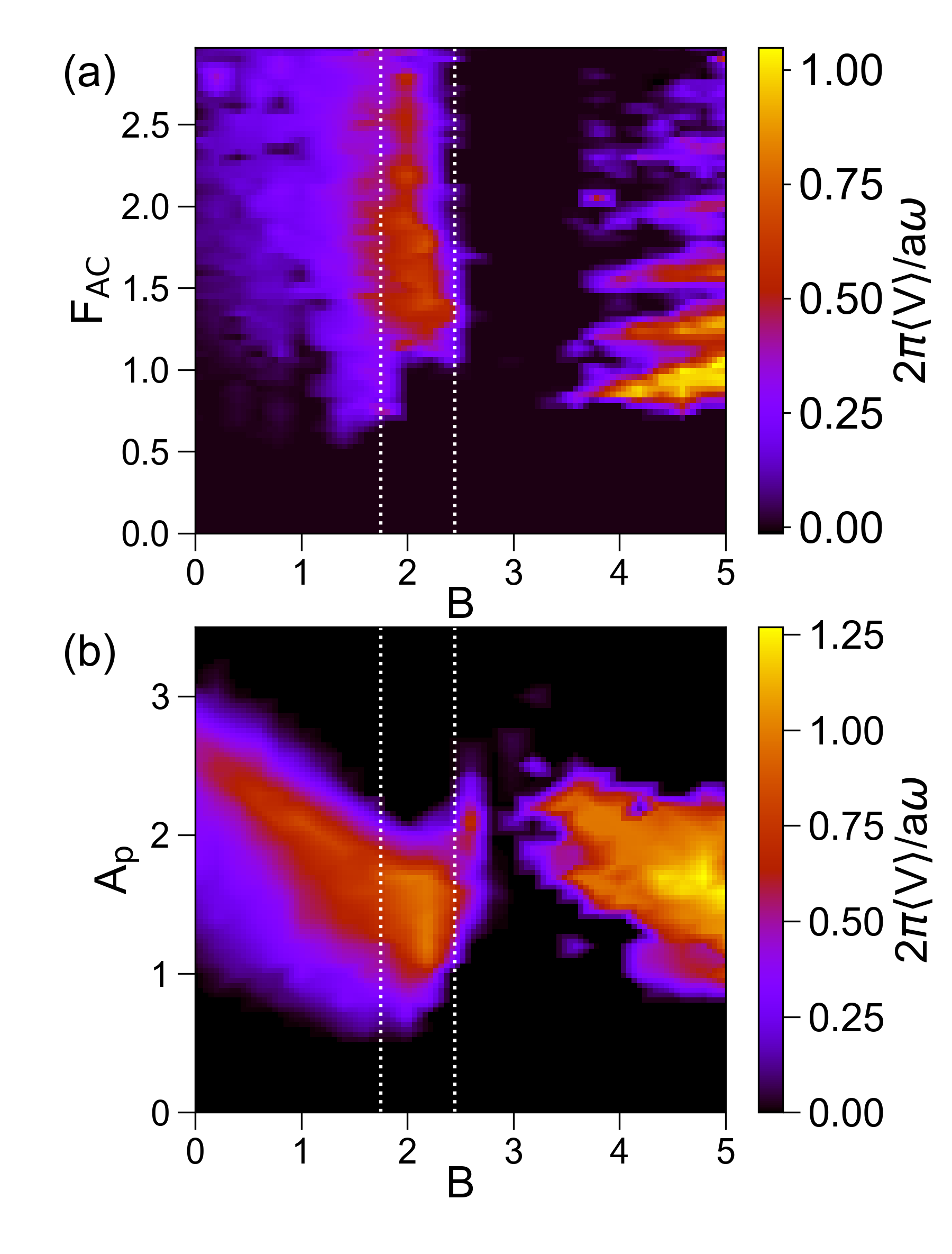}
\caption{
(a) Heat map of $2\pi\langle V\rangle/a \omega$ as a function
  of $F_{\rm AC}$ vs $B$ for a system with $A_{p} = 1.0$.
The dashed lines indicate the separation of the stripe state from the
lower $B$ anisotropic crystal and the higher $B$ bubble states.
(b) The same as a function of $A_{p}$ vs $B$
for fixed $F_{\rm AC} = 1.6$.
In each panel, the ratchet behavior is strongly nonmonotonic and
affected by the morphology of the system, with
stripes and small bubbles showing the strongest ratchet effect.
}
\label{fig:3}
\end{figure}

To more clearly demonstrate the nonmonotonic ratchet effect,
in Fig.~\ref{fig:3}(a)
we plot a heat map of
$2\pi\langle V\rangle/a \omega$ as a function of $F_{\rm AC}$ versus $B$
for a system with $A_{p} = 1.0$ obtained at $F_{\rm AC}$ and $B$ increments
of $0.05$ and $0.2$, respectively.
The dashed lines indicate the separation between the anisotropic crystal
state, the stripe state, and the bubble states.
There is a strong ratchet effect for
$F_{\rm AC} > 1.0$ in 
the stripe phase, while
the anisotropic crystal shows a reduced ratchet effect.
In the large bubble state
for $2.4 < B < 3.8$, the ratchet effect is almost completely absent,
while for $B >3.8$, the small bubble state exhibits
a strong ratchet effect with oscillatory behavior as a function of $F_{\rm AC}$.
In Fig.~\ref{fig:3}(b) we show a heat map of
$2\pi\langle V\rangle/a \omega$  as a function of $A_{p}$ vs $B$
at fixed $F_{\rm AC} = 1.6$.
There is a strong ratchet effect in the stripe and small bubble regimes
and a reduced ratchet effect in the anisotropic crystal and
large bubble regimes.
For large $A_{p}$, the system becomes pinned and the ratchet effect
vanishes since each particle is confined to a single substrate minimum,
while for $F_{p} < 0.6$,
the ratchet effect also disappears because
the system enters a floating phase in which
the particles decouple from the substrate.
At $B = 0.0$, the separation between particles reaches its greatest
extent, and
the pinned state does not appear until
$A_{p} > 3.0$.

\begin{figure}
  \centering
\includegraphics[width=\columnwidth]{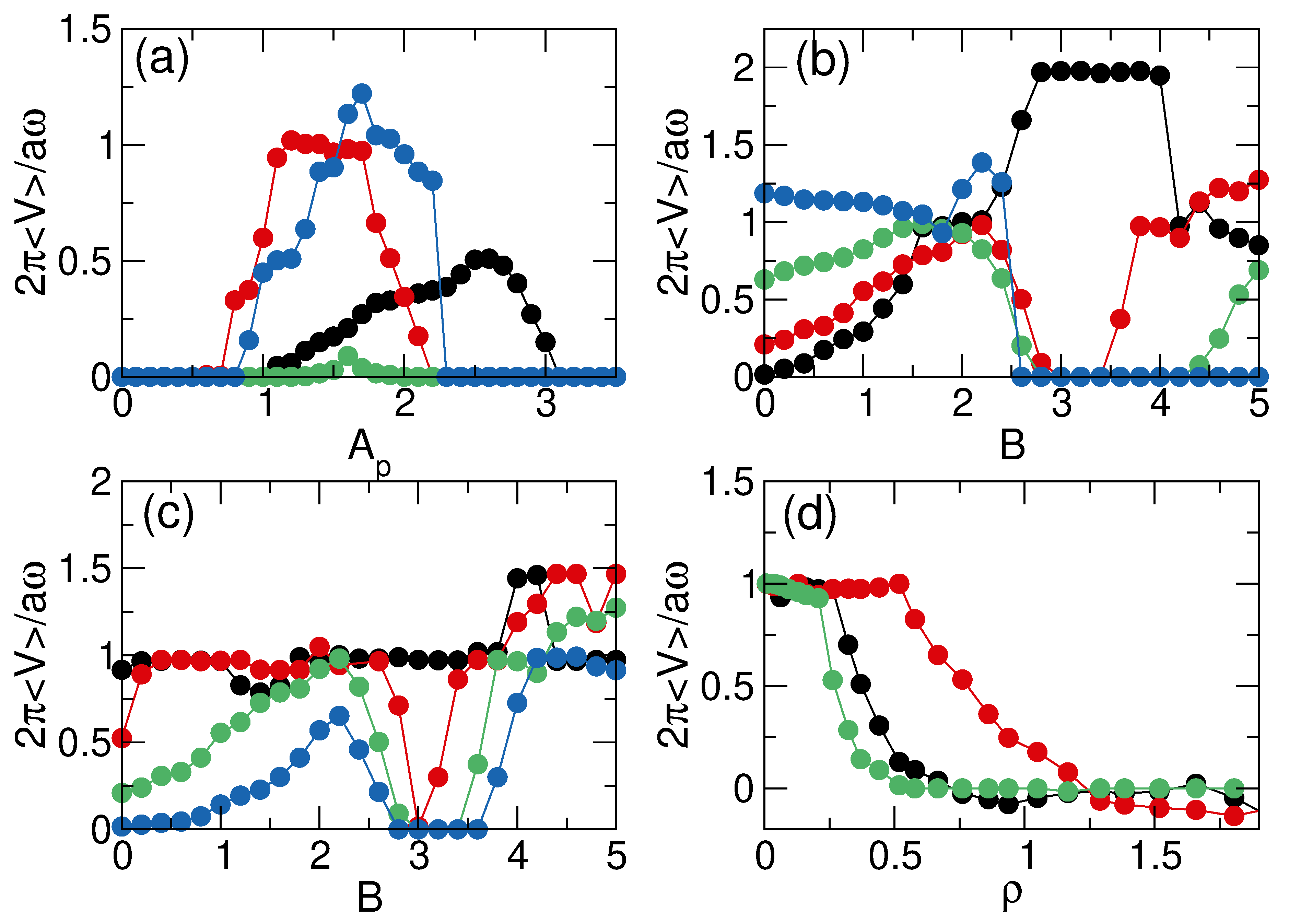}
\caption{(a) $2\pi\langle V\rangle/a\omega$
vs $A_{p}$ for the system in Fig.~\ref{fig:3}(b) with
$\rho=0.44$, $a=4.5$, and $F_{\rm AC}=1.6$
at
$B = 0.0$ (black), 2.2 (red), 2.8 (green) and $4.4$ (blue).        
(b) $2\pi\langle V\rangle/a\omega$
vs $B$ for a system with $\rho = 0.44$,
$A_p=1.6$, and $F_{\rm AC}=1.6$ at different
substrate periods $a = 9.0$ (black),
$4.5$ (red), $3.0$ (green),
and $2.11$ (blue).
(c) $2\pi\langle V\rangle/a\omega$
vs $B$ for a system with $a = 4.5$,
$A_p = 1.6$, and $F_{\rm AC} = 1.6$
at different densities $\rho=0.0925$ (black), $0.2083$ (red),
0.44 (green), and $0.67$ (blue). 
(d) $2\pi\langle V\rangle/a\omega$ vs $\rho $ for a system with
$A_{p} = 1.6$, $F_{\rm AC} = 1.6$, and $a = 4.5$
at $B = 0.04$ in the anisotropic crystal state (black),
$B = 2.2$ in the stripe state (red),
and $B= 2.8$ in the large bubble state (green).
Images of the some of the states appear in the Supplemental Material
\cite{Suppl}.
}
\label{fig:4}
\end{figure}

In Fig.~\ref{fig:4}(a) we plot
$2\pi\langle V\rangle/a\omega$ versus $A_{p}$
for the system in Fig.~\ref{fig:3}(b) at
$B = 0.0$, $2.2$, $2.8$, and $4.4$.
In the large bubble state at $B=2.8$, there is only a
small region of weak ratcheting motion.
To further demonstrate the effects of the morphology
on the ratcheting mechanism,
in Fig.~\ref{fig:4}(b) we plot
$2\pi\langle V\rangle/a\omega$
versus $B$
in samples with $\rho = 0.44$,
$A_p = 1.6$, and $F_{\rm AC} = 1.6$ at different substrate
periods of
$a = 9.0$, $4.5$, $3.0$, and $2.11$.
For $a = 9.0$, a strong ratchet
effect appears for $2.4 < B < 4.0$ since the large bubbles
can easily fit inside the substrate troughs due to the large
spacing between adjacent substrate minima, and are able to travel
a net distance of $2a$ per ac drive cycle.
In contrast, when $a = 2.11$ there is no ratchet
effect for $B > 2.4$
because the bubble widths are much larger than the substrate spacing.
The $a = 3.0$ and $a = 4.5$ curves are similar to each other and show that
stripes and small bubbles ratchet
while large bubbles do not; however, for $a = 3.0$,
the minimum value of $B$ for which the small bubbles begin to ratchet
shifts upward.

In Fig.~\ref{fig:4}(c) we show the $a=4.5$ system from
Fig.~\ref{fig:4}(b) at varied $\rho=0.0925$,
$0.2083$,  $0.44$, and $0.67$.
For $\rho = 0.0925$, the total number of particles in the system is relatively
small, so only small 
anisotropic bubbles form over the entire range of $B$ considered,
and there is little variation in the ratchet effect.
When $\rho = 0.2083$, the system forms anisotropic bubbles,
and there is a dip in the ratchet effect
in the large bubble window that appears at $B= 3.0$.
For $\rho = 0.44$ and $\rho=0.67$, there is a reduced ratchet effect
in the anisotropic crystal phase, a peak
in the ratcheting motion in the stripe phase,
an absence of ratcheting
in the large bubble state, and a return to ratcheting motion
in the small bubble regime.
In Fig.~\ref{fig:4}(d) we plot
$2\pi\langle V\rangle/a\omega$ versus $\rho$
for systems with
$A_{p} = 1.6$, $F_{\rm AC} = 1.6$, and $a = 4.5$
at $B = 0.04$ in the anisotropic crystal regime,
$B = 2.2$ in the stripe regime,
and $B=2.8$ in the large bubble regime.
For $B = 2.8$, a ratchet effect can occur for $\rho < 0.25$ when the bubbles
become small enough, due to the reduced number of particles in the system,
to fit inside individual substrate minima, but
for larger $\rho$ there is no ratchet effect.
When $B = 2.2$, a strong ratchet effect occurs for $\rho < 1.23$
and there are weak reversals of the ratchet motion
for $\rho > 1.2$.
For $B =0.4$, a strong
ratchet effect occurs when $\rho < 0.325$,
and there is also a small ratchet reversal for $\rho > 0.73$.

{\it Summary---} 
We have examined the mesophase or pattern forming systems that
exhibit
crystals, stripes, and various bubble states due to a competition
between short-range attraction and long-range repulsion.
When we couple these patterns to a quasi-one-dimensional asymmetric
substrate and apply ac driving,
we find ratcheting motion with strongly nonmonotonic efficiency
as a function of the mesophase morphology.
The nonmonotonicity is a result of how well or poorly the various
mesophases match with the dimensions of the underlying substrate
potential.
The stripes, which can align with the substrate,
show the strongest ratchet effect.
Anisotropic crystals show a weaker ratchet effect,
and for the large bubbles, ratcheting motion is absent
when the diameter of the bubble is larger than the substrate lattice
constant.
In contrast, small bubbles show a strong ratchet effect
because they can fit easily inside individual substrate minima.
We show that these effects are robust
over a wide range of ac driving amplitudes, substrate strengths, and
short-range attraction strength.
Our results have implications for a broad
array of systems since similar
mesophases appear in a wide variety of soft and hard matter systems,
and numerous methods are available to create modulated substrates
and apply ac driving to such systems.
Since
the different morphologies show different ratchet effectiveness,
our results
could be used to
design protocols for sorting patterns in mesophase-forming systems.

\begin{acknowledgements}
We gratefully acknowledge the support of the U.S. Department of
Energy through the LANL/LDRD program for this work.
This work was supported by the US Department of Energy through
the Los Alamos National Laboratory.  Los Alamos National Laboratory is
operated by Triad National Security, LLC, for the National Nuclear Security
Administration of the U. S. Department of Energy (Contract No. 892333218NCA000001).
\end{acknowledgements}

\bibliography{mybib}
\clearpage


\section*{Supplementary material}

# Supplemental information for "Stripe and Bubble Ratchets on Asymmetric Substrates"

C. Reichhardt and C. J. O. Reichhardt

*Theoretical Division, Los Alamos National Laboratory, Los Alamos, New Mexico 87545 USA*

We show additional images of the particle configurations for states with and without strong ratcheting effects. In Fig. S1 we plot images of the stripe state at $B = 2.2$ from Fig. 4(a) of the main text. When $A_p = 0.2$ in Fig. S1(a), the particles float above the substrate and do not couple to it, resulting in the absence of a ratchet effect, while for $A_p = 3.5$ in Fig. S1(b), the particles remain completely pinned by the substrate and there is no ratchet motion.

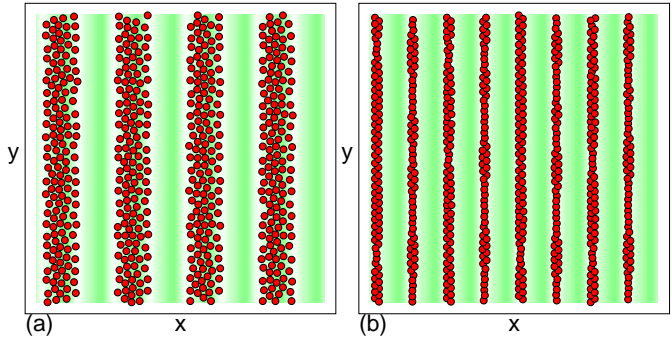

FIG. S1. Image of particle positions (red) and the underlying asymmetric potential (green shading) for samples with $F_{AC} = 1.6$, $a = 4.5$, $\rho = 0.44$, and $B = 2.2$ at (a) $A_p = 0.2$, where there is no ratchet effect and the particles are floating above the substrate, and at (b) $A_p = 3.5$, where the system is pinned.

In Fig. S2(a,b) we plot images of the bubble state from Fig. 4(b) of the main text. At $a = 9.0$ and $B = 2.8$ in Fig. S2(a), the size of the bubbles matches the size of the substrate lattice constant and there is a strong ratchet effect, while at $a = 2.11$ and $B = 2.4$ in Fig. S2(b), the bubbles are too wide to fit fully inside the substrate troughs and there is no ratchet effect. In the anisotropic crystal regime for this sample at $a = 2.11$ and $B = 0.4$, shown in Fig. S2(c), a strong ratchet effect is present.

In Fig. S3 we plot images of the bubble state at $B = 4.4$ from Fig. 4(c) of the main text. For $\rho = 0.0925$ in Fig. S3(a), the small number of particles present in the system produces narrow bubbles that couple well to the substrate and experience a strong ratchet effect. At $\rho = 0.67$ in Fig. S3(b), the bubbles are too wide to fit fully inside the substrate troughs and there is no ratchet effect.

In Fig. S4 we plot images of the from Fig. 4(d) of the main text. For the stripe state at $B = 4.2$ in Fig. S4(a), there is a forward or $+x$ direction ratchet effect. In the

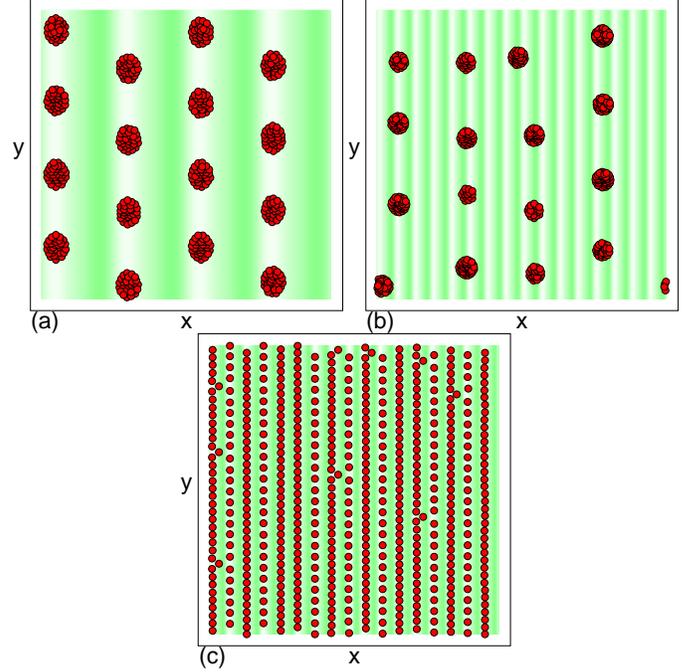

FIG. S2. Image of particle positions (red) and the underlying asymmetric potential (green shading) for samples with $A_p = 1.6$ and $\rho = 0.44$ at (a) $a = 9.0$ and $B = 2.8$, where there is a strong ratchet effect, (b) $a = 2.11$ and $B = 4.4$, where there is no ratchet effect, and (c) $a = 2.11$ and $B = 0.4$, where there is a strong ratchet effect.

bubble state at $B = 4.8$, shown in Fig. S4(b), no ratchet effect appears. Finally, in the anisotropic crystal state at $B = 0.4$ in Fig. S4(c), a weak reverse or $-x$ direction ratchet effect occurs.

**Supplemental Movie 1:** Motion of the system in Fig. 2(a) of the main text with $A_p = 1.6$, $a = 4.5$, $\rho = 0.44$, and $B = 2.2$ at $F_{AC} = 1.6$, where there is a strong ratchet effect.

**Supplemental Movie 2:** Motion of the system in Fig. 2(b) of the main text with $A_p = 1.6$, $a = 4.5$, $\rho = 0.44$, and $B = 2.8$ at $F_{AC} = 1.6$, where there is a weak ratchet effect.

**Supplemental Movie 3:** Motion of the system in Fig. 2(b) of the main text with $A_p = 1.6$, $a = 4.5$, $\rho = 0.44$, and $B = 4.4$ at $F_{AC} = 1.6$, where there is a strong ratchet effect.

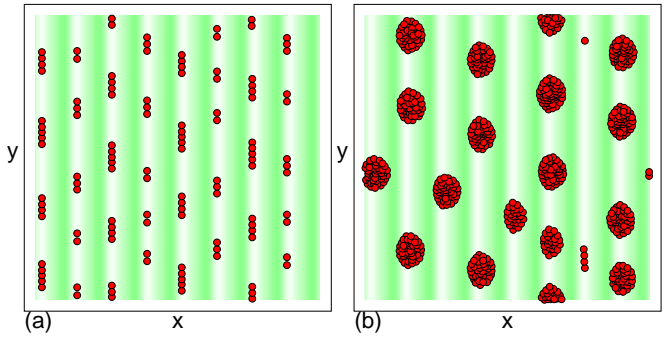

FIG. S3. Image of particle positions (red) and the underlying asymmetric potential (green shading) for samples with $A_p = 1.6$, $a = 4.5$, and $B = 4.4$ at (a) $\rho = 0.0925$, where there is a strong ratchet effect, and (b) $\rho = 0.67$, where there is no ratchet effect.

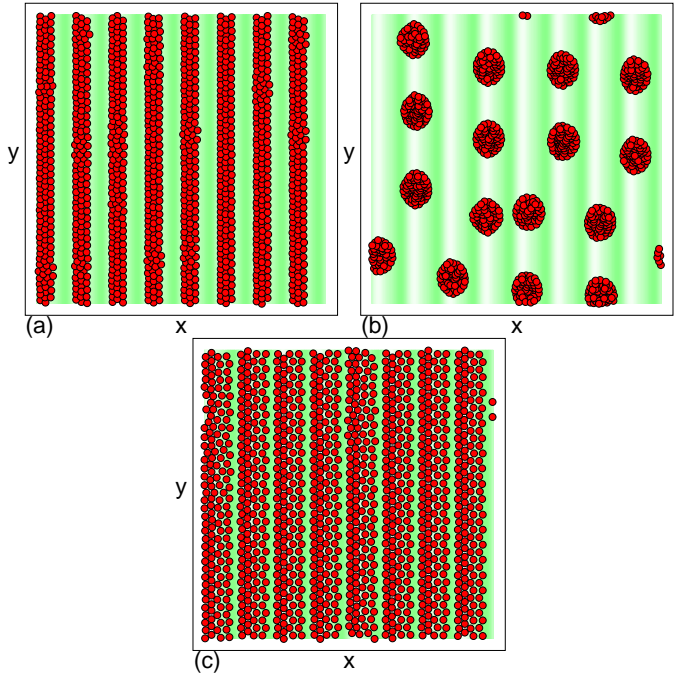

FIG. S4. Image of particle positions (red) and the underlying asymmetric potential (green shading) for samples with $A_p = 1.6$, $a = 4.5$, and $\rho = 0.9382$ at (a) $B = 4.2$, where there is a forward ($+x$ direction) ratchet effect, (b) $B = 4.8$ where there is no ratchet effect, and (c) $B = 0.4$ where there is an anisotropic crystal with a weak reverse ($-x$ direction) ratchet effect.



\end{document}